\documentclass[epj,final]{svjour}

\usepackage{epsfig}
\usepackage{graphicx}

\newcommand{\bmle}{\begin{equation}\begin{array}{rl}}
\newcommand{\emle}{\end{array}\end{equation}}

\newcommand{\be}{\begin{equation}}
\newcommand{\ee}{\end{equation}}
\newcommand{\ba}[1]{\begin{array}{#1} }
\newcommand{\ea}{\end{array}}
\newcommand{\bea}{\begin{eqnarray}}
\newcommand{\eea}{\end{eqnarray}}
\newcommand{\sub}[1]{_{#1}}
\newcommand{\sups}[1]{^{#1}}
\newcommand{\mpi}{m\sub{\pi}}
\newcommand{\fPi}{f\sub{\pi}}
\newcommand{\GE}[1]{G\sub{E}\sups{#1}}
\newcommand{\GM}[1]{G\sub{M}\sups{#1}}
\newcommand{\expval}[1]{\left\langle #1 \right\rangle}
\newcommand{\rexp}{\expval{r^2}}

\begin{document}

\title{Nucleon electromagnetic form factors from lattice QCD}

\author{J.D.~Ashley \and D.B.~Leinweber \and A.W.~Thomas \and R.D.~Young}

\institute{Special Research Centre for the Subatomic Structure of Matter and \\
Department of Physics, University of Adelaide, \\
Adelaide SA 5005 Australia}

\date{Received: date / Revised version: date}

\abstract{
It is imperative that lattice QCD serve to develop our understanding of
hadron structure and, where possible, to guide the interpretation of
experimental data. There is now a great deal of effort directed at the
calculation of the electroweak form factors of the nucleon, where
for example, measurements at Jefferson Laboratory 
have recently revealed surprising
behaviour in the ratio $G_E/G_M$. 
While, for the present, calculations within the framework of lattice QCD
are limited to relatively large quark mass, there has been considerable
progress in our understanding of how to extrapolate to the chiral limit.
Here we report the results of the application of these techniques to the
most recent form factor data from the QCDSF Collaboration. The level of
agreement with all of the form factors, for $Q^2$ below 1 GeV$^2$, is
already impressive. 
\PACS{
      {12.38.Gc}{Lattice QCD calculations }  
	  \and {11.30.Rd}{Chiral symmetries}
      \and {13.40.Gp}{Electromagnetic form factors} 
	  \and {14.20.Dh}{Protons and neutrons}
     } 
} 

\maketitle


\section{Introduction}
\label{sec:intro}
The electromagnetic form factors of the nucleon provide a fundamental
constraint for any theoretical description of the structure of the
nucleon \cite{Thomas:kw}. Even though they have been studied
experimentally for more than 50 years, recent experiments at Jefferson
Lab \cite{Jones:1999rz} have revealed surprising new behaviour in the
ratio $G_E/G_M$ for the proton. In addition, there are a range of other
new results on the neutron electric and magnetic form factors
\cite{Gao:2003ag} -- this is an exciting and rapidly developing field.

In this context it is vital that lattice QCD, our only rigorous method
of solving non-perturbative QCD, be used to inform our understanding of
this data and the various models of hadron structure that are used to
describe it. There was a relatively long hiatus in calculations of
nucleon form factors in lattice QCD after the pioneering work of
Leinweber and collaborators in the early 90's
\cite{Leinweber:1990dv,Leinweber:1991vc,Draper:uu}. However, the
activity has intensified in the last few years 
\cite{Gockeler:2003ay,LATT02,Dong:1998tj}. 
In spite of this activity, limitations in
computer speed mean that we are currently limited to lattice simulations
at quark masses a factor 5-10 higher than the physical values and these
calculations are currently made in quenched approximation (QQCD). 

If one is to compare these state of the art simulations with
experiment, it is necessary to make an extrapolation as a function of
quark mass to the physical region \cite{Thomas:2002sj,Bernard:2002yk}.
This extrapolation is made non-trivial by the non-analytic behaviour
as a function of $m_q$ which follows from the fact that chiral
symmetry is dynamically broken in QCD. Until recently the absence of
data has meant that efforts at chiral extrapolation of form factors
has been focussed on baryon magnetic moments
\cite{Cloet:2003jm,Hackett-Jones:2000qk,Leinweber:1998ej,Hemmert:2002uh}
and charge radii \cite{Hackett-Jones:2000js,Dunne:2001ip}. The focus
on non-analyticity \cite{Leinweber:2001ui} of hadron electromagnetic
properties has inspired considerable investigation of the formal
constraints in both full QCD and QQCD
\cite{Arndt:2003ww,Arndt:2001ye,Savage:2001dy,Beane:2002vq,Chen:2001eg}.

The practical issue is then how to incorporate these formal chiral
constraints into a practical chiral extrapolation, given that the
radius of convergence of the formal expansion dictated by chiral
perturbation theory is rather small \cite{Young:2002ib}. In the case
of the nucleon mass, where there is extremely accurate data in full
QCD from the CP-PACS Collaboration \cite{AliKhan:2001tx}, one can
formally demonstrate the model independence of the extrapolation
procedure \cite{Young:2002ib,Leinweber:1999ig,Leinweber:2003dg}. For
the magnetic moments the data is only now improving to the point where
the model independence of the choice of finite range regulator
\cite{Donoghue:1998bs} can be examined \cite{Cloet:2003jm}. In the
case of the form factors this is not yet possible.

The procedure employed in this paper is
to draw on the earlier phenomenological experience with chiral
extrapolations of the magnetic moments and charge radii, taking simple
phenomenological functional forms which build in the correct
non-analytic behaviour and any other asymptotic constraints that are
known \cite{Panic02}. For the present the accuracy of 
the lattice data is not such that
it could discriminate between different functional forms for the
$Q^2$-dependence of the nucleon form factors. In particular, it cannot
yet address the JLab issue of whether $G_E/G_M$ decreases as $Q^2$
increases. We therefore parametrize separately the isoscalar and
isovector lattice data, at a given value of
$m_q$, or $m_\pi$, as a dipole and then extrapolate the dipole mass as
a function of $m_\pi$, building in the appropriate chiral constraints.
Clearly this simple approach can be systematically improved as the
lattice simulations become capable of making a better 
discrimination between possible functional forms. For the present we
shall see that our relatively simple approach already produces quite
impressive results.

\section{Extrapolations}
\label{sec:extrapolations}
It is well known \cite{Thomas:kw} that the $Q^2$-dependence of the 
nucleon Sachs electromagnetic form factors 
(with the exception of the neutron electric form factor 
$\GE{n}$) are described in first approximation by a dipole form 
\begin{equation}
G(Q^2) = \frac{G(0)}{(1+Q^2/\Lambda^2)^2} .
\label{eq:dipole}
\end{equation}
Here $\Lambda$ is the dipole mass and the charge form factor of the
proton satisfies $\GE{p}(0)=1$, while 
$\GM{p}(0)=\mu\sub{p}$ and $\GM{n}(0)=\mu\sub{n}$ are the 
proton and neutron magnetic moments. As explained in the Introduction, we
will use this phenomenological fact to construct a simple but effective 
extrapolation formula.

To isolate the chiral behaviour of the form factors we rearrange them into 
isovector and isoscalar combinations:
\be
G^{v} = G^{p} - G^{n}
\label{eq:Gvdefn}
\ee
and
\be
G^{s} = G^{p} + G^{n}
\label{eq:Gsdefn} ,
\ee
respectively. These also display a dipole-like $Q^2$-dependence but with 
different dipole masses and magnetic moments.

To extrapolate lattice QCD results for the electromagnetic form factors 
from the large pion masses at which they are calculated to the physical regime, 
we extract dipole masses and magnetic moments from the lattice data and then 
extrapolate these as a function of $m_\pi$.
\begin{figure}[ht]
\begin{center}
{\epsfig{file=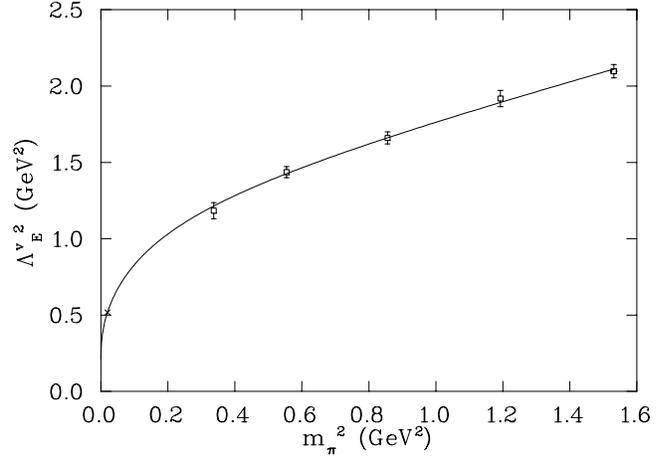, width=6cm, angle=90}}
\caption{Fit to values of the isovector electric form factor dipole mass
extracted from lattice data with lattice spacing $a=0.051$~fm. The
physical value predicted by the fit is also indicated.}
\label{fig:Lambda2 GEv 3}
\end{center}
\end{figure}
\begin{figure}[ht]
\begin{center}
{\epsfig{file=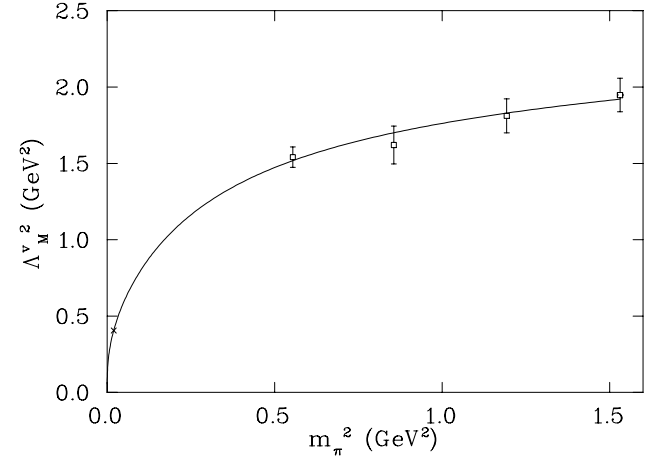, width=6cm, angle=90}}
\caption{Fit to values of the isovector magnetic form factor dipole mass
extracted from lattice data with lattice spacing $a=0.051$~fm. The
physical value predicted by the fit is also indicated.}
\label{fig:Lambda2 GMv 3}
\end{center}
\end{figure}

{}Following Refs.~\cite{Hackett-Jones:2000qk,Leinweber:1998ej}, one
can use a Pad\'e approximant which builds in both the correct chiral
non-analytic behaviour as $m_\pi \rightarrow 0$ and the correct
asymptotic behaviour as $m_q \rightarrow \infty$ to extrapolate the
neutron and proton magnetic moments
\be
\label{eq:Pade}
     \mu_{i}(\mpi) = \frac{\mu_0}{1 - \frac{\chi\sub{i}}{\mu_0}\mpi 
     + c \mpi^2 }.
\ee
The chiral coefficients for the isovector and isoscalar moments are
$\chi_{v} = -8.82$ and $\chi_{s} = 0$ respectively
and $\mu_0$ and $c$ are fitting parameters, to be determined by the
lattice data.

In order to build a suitable extrapolating function for the
$Q^2$-dependence of the form factors, we use the connection between the
mass parameter in a dipole form factor and the corresponding mean-square
radius. For the isovector magnetic form factor the mean square radius is
\be
\label{eq:dGdQ2}
\langle r^2 \rangle_{M}^v = -\frac{6}{\GM{v}(0)}\frac{d\GM{v}}{dQ^2} \, 
|_{Q^2 = 0} \, .
\ee
Comparing this to the dipole of Eq.~(\ref{eq:dipole}), we find an expression 
relating the dipole mass to $\rexp\sub{M}^v$,
\be
\label{eq:Lambda2 r2}
    (\Lambda\sub{M}\sups{v})^2 = \frac{12}{\rexp\sub{M}^v}.
\ee
The chiral behaviour of the magnetic mean squared radius is known from chiral 
perturbation theory \cite{Bernard:1995dp}
\be
\label{eq:rM chiral1}
    \rexp\sub{M}^v \;\sim\; \frac{\chi_1}{\mpi} +
    \chi_2\ln(\frac{\mpi}{\mu}).
\ee
The constants $\chi_1$ and $\chi_2$ are given by
\bea
\label{chiral consts}
    \chi_1 &=& \frac{g\sub{A}^2 m\sub{N}}{8\pi \fPi^2\kappa\sub{v}} \, ,
    \label{eq:chi1}\\
    \chi_2 &=& -\frac{5g\sub{A}^2+1}{8\pi^2 \fPi^2}\label{eq:chi2} \, ,
\eea
where $g\sub{A}=1.27$ is the axial coupling constant and 
$\fPi=93$~MeV is the pion decay constant, $m\sub{N}=940$~MeV
is the nucleon mass and $\kappa\sub{v}\approx 4.2$ is the isovector anomalous
magnetic moment of the nucleon (in the chiral limit). 
\begin{figure}[ht]
\begin{center}
{\epsfig{file=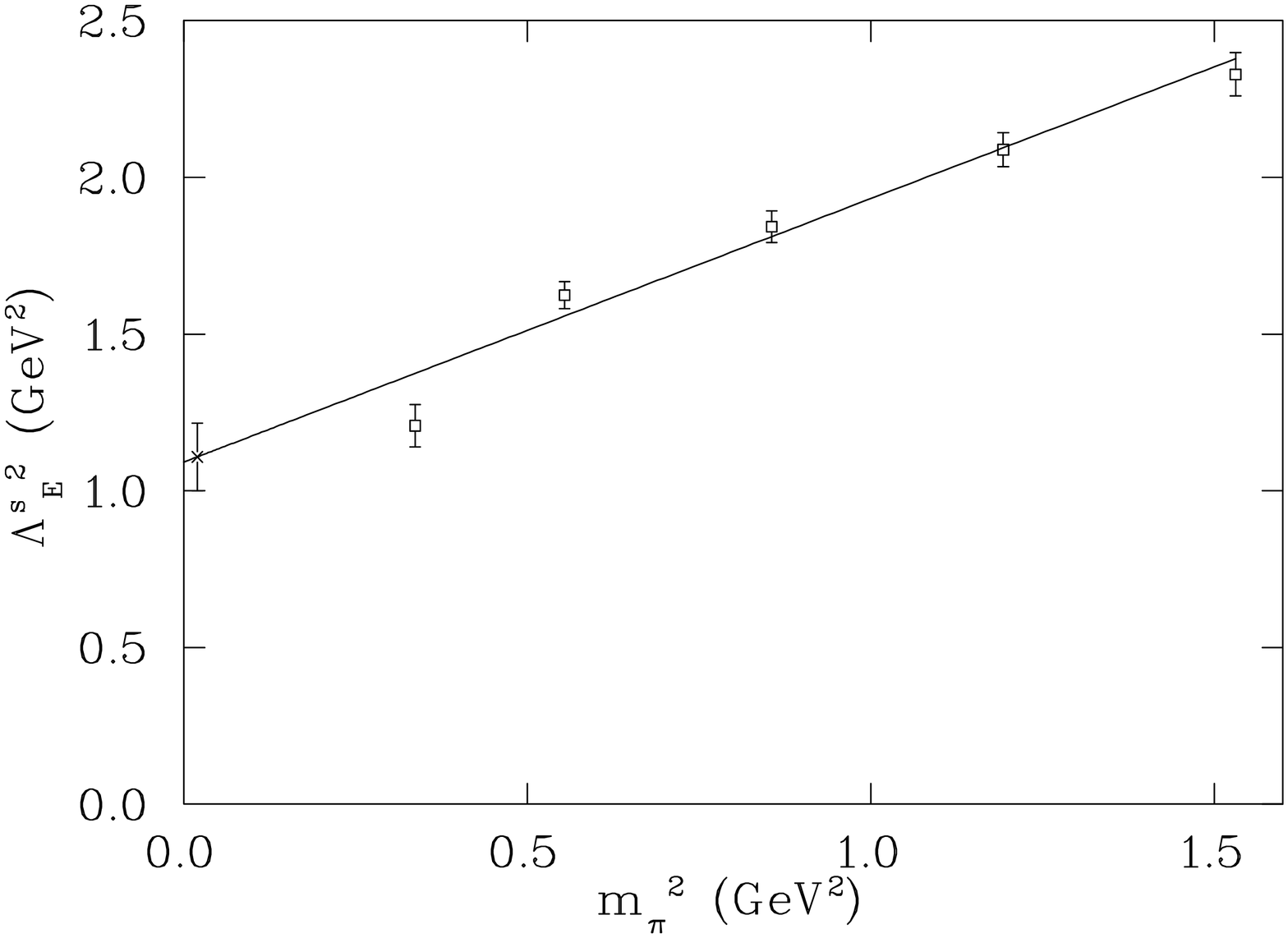, width=6cm, angle=90}}
\caption{Linear fit to values of the isoscalar electric form factor
dipole mass
extracted from lattice data with lattice spacing $a=0.051$~fm. The
physical value predicted by the fit is also indicated.}
\label{fig:Lambda2 GEs 3}
\end{center}
\end{figure}
\begin{figure}[ht]
\begin{center}
{\epsfig{file=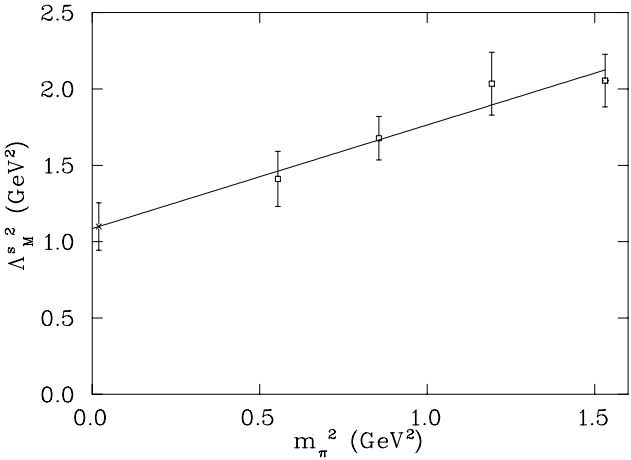, width=6cm, angle=90}}
\caption{Linear fit to values of the isoscalar magnetic form factor
dipole mass
extracted from lattice data with lattice spacing $a=0.051$~fm. The
physical value predicted by the fit is also indicated.}
\label{fig:Lambda2 GMs 3}
\end{center}
\end{figure}

Earlier experience with nucleon properties as a function of quark mass
has suggested that, as a consequence of the finite size of the source of
the pion field, chiral loops are strongly suppressed for $m_\pi > 0.4$
GeV \cite{Detmold:2001hq}. Thus in order to build an effective
chiral extrapolation formula one needs to modify the non-analytic terms
given above so that they are suppressed above this mass.
{}For guidance as to an appropriate parametrization of this suppression
we have evaluated the pion loops that give rise to the non-analytic chiral terms
using the Cloudy Bag Model (CBM) 
\cite{Thomas:1982kv,Theberge:1981pu,Lu:1997sd}.
The results led us to replace Eq.~(\ref{eq:rM chiral1}) with the expression
\be
\label{eq:rM chiral2}
    \rexp\sub{M}^v \;\sim\; 
    \frac{\chi_1}{\mpi}\frac{2}{\pi}\arctan(\mu/\mpi)
    + \frac{\chi_2}{2}\ln(\frac{\mpi^2}{\mpi^2 + \mu^2}) \, ,
\ee
which ensures the correct chiral behaviour at low-$\mpi$ but suppresses it
at values larger than $\mu$.
\begin{figure}[ht]
\begin{center}
{\epsfig{file=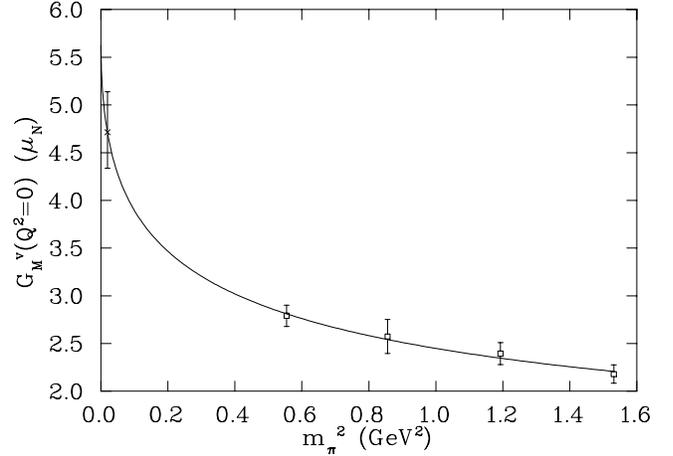, width=6cm, angle=90}}
\caption{Fit to values of the isovector magnetic moment
extracted from lattice data with lattice spacing $a=0.051$~fm. The
physical value predicted by the fit is also indicated.}
\label{fig:GMv0 3}
\end{center}
\end{figure}
\begin{figure}[ht]
\begin{center}
{\epsfig{file=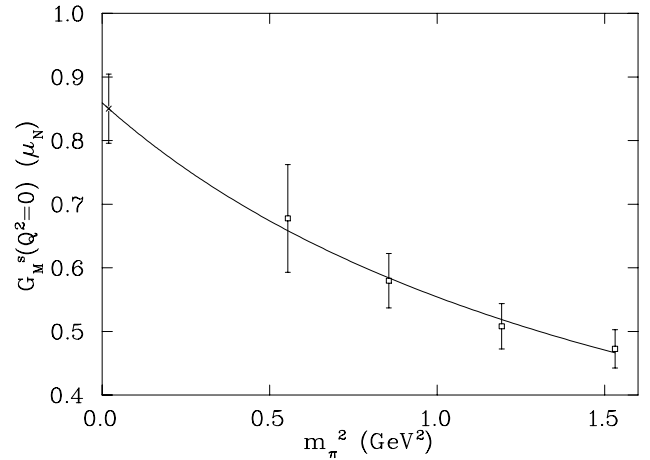, width=6cm, angle=90}}
\caption{Fit to values of the isoscalar magnetic moment
extracted from lattice data with lattice spacing $a=0.051$~fm. The
physical value predicted by the fit is also indicated.}
\label{fig:GMs0 3}
\end{center}
\end{figure}

Substituting this functional form into Eq.~(\ref{eq:Lambda2 r2}) and 
introducing a linear dependence on $\mpi^2$ to give the expected  
behaviour at large quark mass, we get the following expression for 
the dipole mass associated with the isovector magnetic form factor
\be
\label{eq:Lambda2 GMv}
    (\Lambda\sub{M}\sups{v})^2 = \frac{12(1+A_1\mpi^2)}
    {A_0 + \frac{\chi_1}{\mpi}\frac{2}{\pi}\arctan(\mu/\mpi) +
    \frac{\chi_2}{2}\ln\Bigl(\frac{\mpi^2}{\mpi^2+\mu^2}\Bigr)} \, .
\ee
Here $A_0$ and $A_1$ are unknown parameters, adjusted to fit the lattice
data.

We can perform a similar analysis to find an expression for 
the isovector electric 
form factor dipole mass $\Lambda\sub{E}\sups{v}$, combining the non-analytic 
chiral behaviour predicted by $\chi$PT at low $\mpi^2$ with the slowly varying 
behaviour observed in lattice QCD at large masses. 
Using the equivalent of 
Eq.~(\ref{eq:dGdQ2}) for the electric form factor, we can relate the dipole 
mass to the mean charge radius by 
\be
\label{eq:Lambda2 r2 E}
    (\Lambda\sub{E}\sups{v})^2 = \frac{12}{\rexp\sub{E}^v}.
\ee
The chiral behaviour of the isovector mean charge radius is like that of the 
magnetic radius but with no $1/\mpi$ term \cite{Bernard:1995dp},
\be
\label{eq:rE chiral1}
    \rexp\sub{E}^v  \sim \chi_2 \ln(\frac{\mpi}{\mu}) \, .
\ee
Combining this chiral behavior with a linear
dependence on $\mpi^2$ produces
the following predicted form for the isovector electric form factor dipole
mass,
\be
\label{eq:Lambda2 GEv}
    (\Lambda\sub{E}\sups{v})^2 = \frac{12(1+B_1\mpi^2)}{B_0 + 
    \frac{\chi_2}{2} \ln(\frac{\mpi^2}{\mpi^2+\mu^2})} \, ,
\ee
where $B_0$ and $B_1$ are general fitting parameters.

Having isolated the chiral non-analytic behaviour to the 
isovector form factors, we expect 
that the isoscalar dipole masses should be 
roughly linear in $\mpi^2$ -- as observed in 
CBM results.
\begin{figure}[ht]
\begin{center}
{\epsfig{file=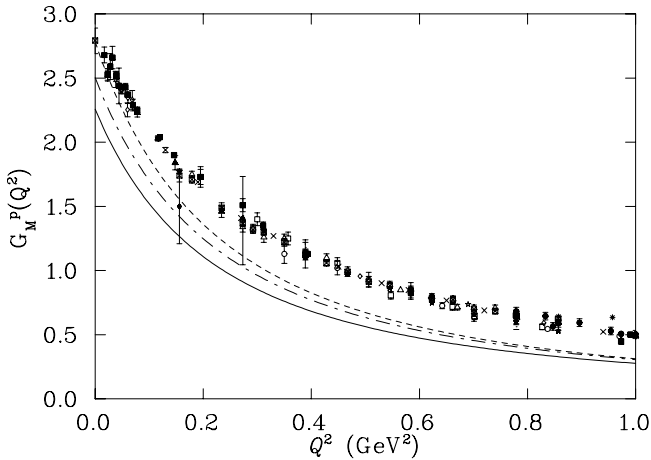, width=6cm, angle=90}}
\caption{The magnetic form factor for the proton extrapolated from
lattice data
with
lattice spacings $a=0.093$~fm (single line), $a=0.068$~fm (dash-dot line)
and $a=0
.051$~fm (dotted line)
and compared to experimental results.}
\label{fig:GMp all}
\end{center}
\end{figure}
\begin{figure}[ht]
\begin{center}
{\epsfig{file=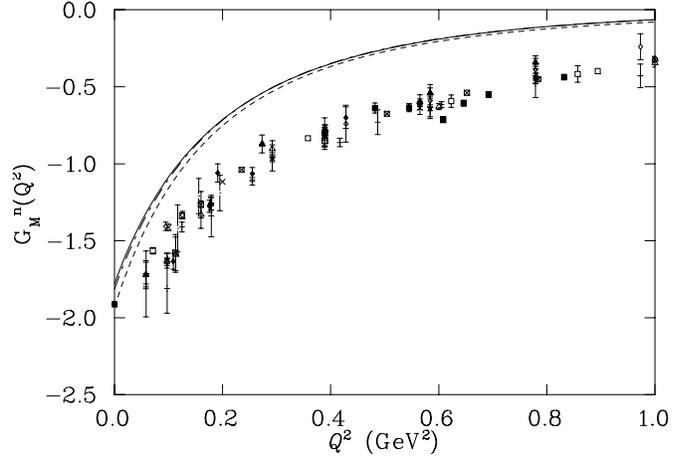, width=6cm, angle=90}}
\caption{The magnetic form factor for the neutron extrapolated from
lattice data
with lattice spacings $a=0.093$~fm (single line), $a=0.068$~fm (dash-dot line)
and $a=0.051$~fm (dotted line)
and compared to experimental results.}
\label{fig:GMn all}
\end{center}
\end{figure}

The forms chosen to represent the dipole masses as a function of $m_\pi$
are able to reproduce the predictions 
of the CBM \cite{Lu:1997sd} very well.

\section{Results}
\label{sec:results}
The QCDSF Collaboration \cite{Gockeler:2003ay} has recently reported 
QQCD results for the isovector and isoscalar nucleon  
electric and magnetic form factors calculated at different $Q^2$ 
values and pion masses and for three different lattice spacings 
($\beta = 6.0, \,  6.2$ and 6.4). Using this data, with the
scale set using the Sommer method ($r_0 = 0.5$fm \cite{Guagnelli:1998ud}), 
we have plotted the form factors 
$\GM{v}$, $\GM{s}$, $\GE{v}$ and $\GE{s}$, calculated on the lattice as
a function of $Q^2$, at each pion mass  
and lattice spacing, and fit each graph with a dipole form, finding a best-fit
dipole mass and magnetic moment.
Plots of these best-fit dipole masses and magnetic moments against $\mpi^2$ 
{}for the lattice spacing $a=0.051$~fm are shown 
in Figs. \ref{fig:Lambda2 GEv 3}, 
\ref{fig:Lambda2 GMv 3}, \ref{fig:Lambda2 GEs 3} and \ref{fig:Lambda2 GMs 3}.
\begin{figure}[ht]
\begin{center}
{\epsfig{file=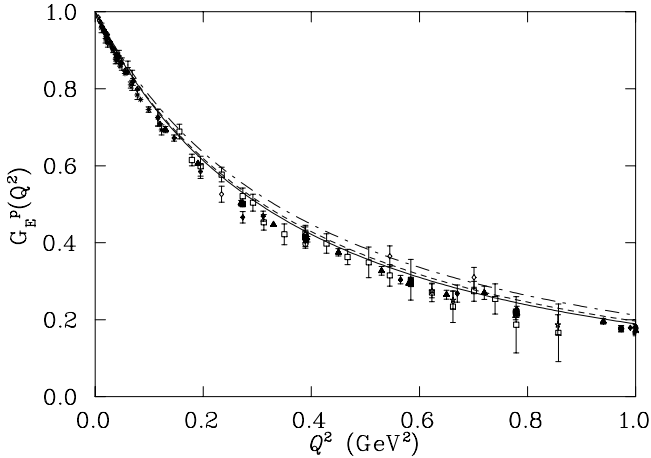, width=6cm, angle=90}}
\caption{The electric form factor for the proton extrapolated from
lattice data
with
lattice spacings $a=0.093$~fm (single line), $a=0.068$~fm (dash-dot line)
and $a=0
.051$~fm (dotted line)
and compared to experimental results.}
\label{fig:GEp all}
\end{center}
\end{figure}
\begin{figure}[ht]
\begin{center}
{\epsfig{file=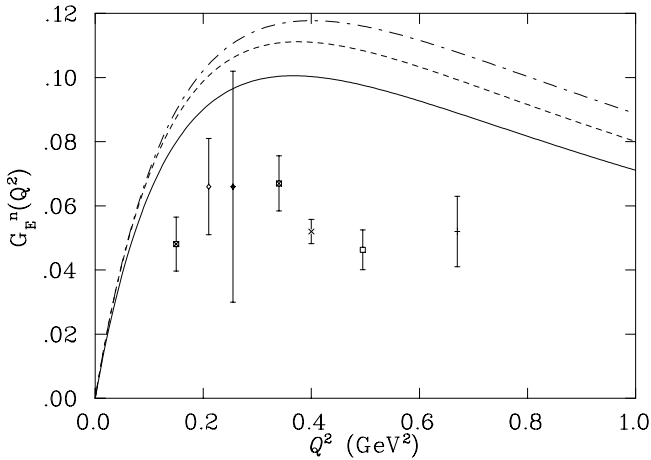, width=6cm, angle=90}}
\caption{The electric form factor for the neutron extrapolated from
lattice data
with lattice spacings $a=0.093$~fm (single line), $a=0.068$~fm (dash-dot
line)
and $a=0.051$~fm (dotted line)
compared to experimental results.}
\label{fig:GEn all}
\end{center}
\end{figure}

The electric and magnetic isovector dipole masses are fitted with
functions of the form Eq.~(\ref{eq:Lambda2 GMv}) and
Eq.~(\ref{eq:Lambda2 GEv}), respectively. The scale $\mu=0.41$~GeV was
chosen to give the best simultaneous fit to the lattice data for both
the electric and magnetic isovector form factors at all three lattice
spacings.  We observe that the curves fit the data well and predict
physical results significantly different from those obtained by a
naive linear extrapolation.  In both cases, the form of the fitting
functions greatly restricts the physical value, resulting in a small
error. Finally, we note that the electric and magnetic isoscalar
dipole mass plots are fitted with linear functions in $\mpi^2$,
because of the absence of any leading non-analytic behaviour in this
case.

The magnetic moment plots are extrapolated using the Pad\'e 
approximant, Eq.~(\ref{eq:Pade}). The resulting errors in the physical magnetic 
moments dominate over errors in the dipole mass and are the main 
source of error in our results for the physical magnetic form factors.
Similar fits were produced for lattice results at lattice spacings $a=0.068$~fm 
and $a=0.093$~fm.

Using the extrapolated physical values for the dipole masses and magnetic 
moments, we reconstruct the electric and magnetic form factors for the proton
and neutron as a sum of two dipoles in $Q^2$. Figs.~\ref{fig:GMp all}, 
\ref{fig:GMn all}, \ref{fig:GEp all} and \ref{fig:GEn all}
show the electric and magnetic form factors for the proton and 
neutron extrapolated from that lattice data at 
each of the three lattice spacings 
($a=0.093$~fm, $a=0.068$~fm and $a=0.051$~fm) compared to 
the experimental data from Refs. \cite{DATA}.

Except for the case of the neutron electric form factor 
(which is very sensitive because it is the result of a cancellation
between $\GE{v}$ and $\GE{s}$ which yields a relatively small result)
our extrapolated curves 
match the experimental points very well, particularly for 
$G_E^p$ and $G_M^p$.  
Although our curve for $\GE{n}$ is almost twice as big as 
the experimental values, it does describe the
correct shape but peaks at slightly too high a value of $Q^2$.

\section{Conclusion}
\label{sec:conclusion}
We have proposed a relatively simple approach to the extrapolation of
lattice QCD data for the nucleon electromagnetic form factors. The data
from QCDSF has been parametrized by a simple dipole form, with the
dipole mass parameter taken to be a function of $m_\pi$ which
incorporates the leading non-analytic behaviour of chiral perturbation
theory. For the isoscalar case, where there is no leading non-analytic
behaviour, the extrapolation is a simple linear function of $m_\pi^2$,
while for the isovector case we have used a functional form suggested by
studies based on the cloudy bag model, which suppress the chiral
behaviour for pion masses larger than 400 MeV. 

The level of agreement between the empirical $Q^2$ dependence of the proton
electric form factor and the proton and neutron magnetic form factors is
impressive. This is quite remarkable when one considers that the data is
based on quenched approximation and is extrapolated from a light quark mass
a factor of twenty above the physical mass. This is only possible
because of the remarkable fact that pion loops are suppressed by finite
size effects once the pion Compton wavelength is smaller than the size
of the source -- empirically, $m_\pi > 0.4$ GeV. The extrapolation of the
data has, of course, been performed using chiral coefficients
appropriate to full QCD.

One of the main open questions in the analysis concerns the neutron electric
form factor, which will always be a challenge given that it vanishes at
$Q^2=0$ -- as a result of the cancellation between the isoscalar and
isovector contributions. In addition, we observe that there appears to
be some residual $\beta$ dependence in the value of the nucleon magnetic
moments and this is the major ambiguity in the current reconstruction of
the proton and neutron magnetic form factors. It is encouraging that the
results for $\beta = 6.4$, corresponding to the smallest lattice spacing,
is in the best agreement with the experimental data. Future lattice simulations
will undoubtedly clarify this issue and also give us data at lower
values of the quark mass. It would also be valuable to have full QCD
simulations (rather than QQCD) to ensure that there is no unanticipated
systematic difference. Most importantly, as the accuracy of the data and
the range of $Q^2$ covered increases, we can extend the present analysis
by employing more complicated fitting functions which should allow us to
test the behaviour of properties such as $G_E/G_M$ against experiment.

\section*{Acknowledgements}
We would like to thank G. Schierholz for helpful communications
concerning the lattice data from QCDSF and J. Kelly, R. Madey and A.
Semenov for providing a compilation of the experimental data on the
nucleon form factors. This work was supported by the
Australian Research Council and the University of Adelaide.


%

\begin{thebibliography}{}
%
\bibitem{Thomas:kw}
A.~W.~Thomas and W.~Weise,
``The Structure Of The Nucleon,'' Wiley-VCH (Berlin, 2001) pp.389 
%
\bibitem{Jones:1999rz}
M.~K.~Jones {\it et al.}  [Jefferson Lab Hall A Collaboration],
Phys.\ Rev.\ Lett.\  {\bf 84} (2000) 1398
[arXiv:nucl-ex/9910005].
%
\bibitem{Gao:2003ag}
H.~y.~Gao,
Int.\ J.\ Mod.\ Phys.\ E {\bf 12} (2003) 1
[arXiv:nucl-ex/0301002].
%
\bibitem{Leinweber:1990dv}
D.~B.~Leinweber, R.~M.~Woloshyn and T.~Draper,
Phys.\ Rev.\ D {\bf 43} (1991) 1659.
%
\bibitem{Leinweber:1991vc}
D.~B.~Leinweber,
Phys.\ Rev.\ D {\bf 45} (1992) 252.
%
\bibitem{Draper:uu}
T.~Draper, K.~F.~Liu, D.~B.~Leinweber and R.~M.~Woloshyn,
Nucl.\ Phys.\ A {\bf 527} (1991) 531C.
%
\bibitem{Gockeler:2003ay}
M.~Gockeler, T.~R.~Hemmert, R.~Horsley, D.~Pleiter, P.~E.~Rakow,
A.~Schafer and G.~Schierholz
[QCDSF Collaboration],
arXiv:hep-lat/0303019.
%
\bibitem{LATT02}
Proceedings of the XXth Int. Symposium on Lattice Field Theory,
eds. R. Edwards, J. Negele and D. Richards, Nucl. Phys. 
B(Proc. Suppl.)119 (2003).
%
\bibitem{Dong:1998tj}
S.~J.~Dong, K.~F.~Liu and A.~G.~Williams,
Workshop on Future Directions in Quark Nuclear
Physics, eds. A.W. Thomas and K. Tsushima (World Scientific, Singapore,
1999) p.131.
%
\bibitem{Thomas:2002sj}
A.~W.~Thomas, Nucl. Phys. B(Proc. Suppl.)119 (2003) 50.
arXiv:hep-lat/0208023.
%
\bibitem{Bernard:2002yk}
C.~Bernard {\it et al.}, 
Nucl. Phys. B(Proc. Suppl.)119 (2003) 170.
arXiv:hep-lat/0209086.
%
\bibitem{Cloet:2003jm}
I.~C.~Cloet, D.~B.~Leinweber and A.~W.~Thomas,
Phys.\ Lett.\ B {\bf 563} (2003) 157
[arXiv:hep-lat/0302008].
%
\bibitem{Hackett-Jones:2000qk}
E.~J.~Hackett-Jones, D.~B.~Leinweber and A.~W.~Thomas,
Phys.\ Lett.\ B {\bf 489} (2000) 143
[arXiv:hep-lat/0004006].
%
\bibitem{Leinweber:1998ej}
D.~B.~Leinweber, D.~H.~Lu and A.~W.~Thomas,
Phys.\ Rev.\ D {\bf 60} (1999) 034014
[arXiv:hep-lat/9810005].
%
\bibitem{Hemmert:2002uh}
T.~R.~Hemmert and W.~Weise,
Eur.\ Phys.\ J.\ A {\bf 15}, 487 (2002)
[arXiv:hep-lat/0204005].
%
\bibitem{Hackett-Jones:2000js}
E.~J.~Hackett-Jones, D.~B.~Leinweber and A.~W.~Thomas,
Phys.\ Lett.\ B {\bf 494} (2000) 89
[arXiv:hep-lat/0008018].
%
\bibitem{Dunne:2001ip}
G.~V.~Dunne, A.~W.~Thomas and S.~V.~Wright,
Phys.\ Lett.\ B {\bf 531} (2002) 77
[arXiv:hep-th/0110155].
%
\bibitem{Leinweber:2001ui}
D.~B.~Leinweber, A.~W.~Thomas and R.~D.~Young,
Phys.\ Rev.\ Lett.\  {\bf 86} (2001) 5011
[arXiv:hep-ph/0101211].
%
\bibitem{Arndt:2003ww}
D.~Arndt and B.~C.~Tiburzi,
arXiv:hep-lat/0307003.
%
\bibitem{Arndt:2001ye}
D.~Arndt and M.~J.~Savage,
Nucl.\ Phys.\ A {\bf 697} (2002) 429
[arXiv:nucl-th/0105045].
%
\bibitem{Savage:2001dy}
M.~J.~Savage,
Nucl.\ Phys.\ A {\bf 700} (2002) 359
[arXiv:nucl-th/0107038].
%
\bibitem{Beane:2002vq}
S.~R.~Beane and M.~J.~Savage,
Nucl.\ Phys.\ A {\bf 709} (2002) 319
[arXiv:hep-lat/0203003].
%
\bibitem{Chen:2001eg}
J.~W.~Chen and X.~d.~Ji,
Phys.\ Lett.\ B {\bf 523} (2001) 107
[arXiv:hep-ph/0105197].
%
\bibitem{Young:2002ib}
R.~D.~Young, D.~B.~Leinweber and A.~W.~Thomas,
Prog.\ Part.\ Nucl.\ Phys.\  {\bf 50} (2003) 399
[arXiv:hep-lat/0212031].
%
\bibitem{AliKhan:2001tx}
A.~Ali Khan {\it et al.}  [CP-PACS Collaboration],
Phys.\ Rev.\ D {\bf 65} (2002) 054505
[Erratum-ibid.\ D {\bf 67} (2003) 059901]
[arXiv:hep-lat/0105015].
%
\bibitem{Leinweber:1999ig}
D.~B.~Leinweber, A.~W.~Thomas, K.~Tsushima and S.~V.~Wright,
Phys.\ Rev.\ D {\bf 61} (2000) 074502
[arXiv:hep-lat/9906027].
%
\bibitem{Leinweber:2003dg}
D.~B.~Leinweber, A.~W.~Thomas and R.~D.~Young,
arXiv:hep-lat/0302020.
%
\bibitem{Donoghue:1998bs}
J.~F.~Donoghue, B.~R.~Holstein and B.~Borasoy,
Phys.\ Rev.\ D {\bf 59} (1999) 036002
[arXiv:hep-ph/9804281].
%
\bibitem{Panic02}
A.~W.~Thomas, J.~D.~Ashley, W.~Detmold, D.~B.~Leinweber, W.~Melnitchouk
and R.~D.~Young, Nucl.\ Phys.\ A {\bf 721} (2003) 915.
%
\bibitem{Bernard:1995dp}
V.~Bernard, N.~Kaiser and U.~G.~Meissner,
Int.\ J.\ Mod.\ Phys.\ E {\bf 4} (1995) 193
[arXiv:hep-ph/9501384].
%
\bibitem{Detmold:2001hq}
W.~Detmold, D.~B.~Leinweber, W.~Melnitchouk, A.~W.~Thomas and
S.~V.~Wright,
Pramana {\bf 57} (2001) 251
[arXiv:nucl-th/0104043].
%

\bibitem{Thomas:1982kv}
A.~W.~Thomas,
Physics,''
Adv.\ Nucl.\ Phys.\  {\bf 13} (1984) 1.
%
\bibitem{Theberge:1981pu}
S.~Theberge and A.~W.~Thomas,
Phys.\ Rev.\ D {\bf 25} (1982) 284.
%
\bibitem{Lu:1997sd}
D.~H.~Lu, A.~W.~Thomas and A.~G.~Williams,
Phys.\ Rev.\ C {\bf 57} (1998) 2628
[arXiv:nucl-th/9706019].
%
\bibitem{Guagnelli:1998ud}
M.~Guagnelli, R.~Sommer and H.~Wittig  [ALPHA collaboration],
Nucl.\ Phys.\ B {\bf 535} (1998) 389
[arXiv:hep-lat/9806005].
%
\bibitem{DATA}
H.~Zhu {\it et al.}  [E93026 Collaboration],
Phys.\ Rev.\ Lett.\  {\bf 87} (2001) 081801
[arXiv:nucl-ex/0105001];
%
W.~Xu {\it et al.},
Phys.\ Rev.\ Lett.\  {\bf 85} (2000) 2900
[arXiv:nucl-ex/0008003]; and a compilation of earlier data provided by
J. Kelly, R. Madey and A. Semenov.
%
\end{thebibliography}
\end{document}